# Mesoscale Cyclonic Eddies in the Black Sea Region

V. V. Efimov, M. V. Shokurov, D. A. Yarovaya , and D. Hein
Marine Hydrophysical Institute, National Academy of Sciences of Ukraine, Kapitanskaya ul. 2, Sevastopol, 99011 Ukraine
e-mail: efimov@alpha.mhi.iuf.net

**Abstract**
Results of regional climate modeling PRECIS with high spatial resolution (25 km) were used to investigate mesoscale features of atmospheric circulation in the Black Sea Region for 30-yr period. Method based on Ocubo – Weiss criterion was used to detect and track subsynoptic eddies. Several types of cyclonic eddy were discovered and studied: winter Caucasian coastal, summer Caucasian off-shore, ambient cyclonic eddies, and rare quasitropical cyclones. For Caucasian eddies statistics of their life-time and intensity, as well as histograms of diurnal and seasonal cycles, are presented.

# Introduction

Atmospheric circulation in the Black Sea Region is mostly influenced by large-scale synoptic cyclones which originate in the North Atlantic and move with the westerlies over Europe or the Mideterranen Sea towards the Black Sea. Such cyclones have typical horizontal scale of 2-3 thousand km and time scale of 2-3 days. They mostly occur in winter and their storm tracks are greatly influenced by North-Atlantic Oscillation. Statistics of their storm tracks is well-studied [1]. Such synoptic-scale cyclones are main extratropical synoptic processes and are ambient events for the Black Sea Region.

At the same time, orographic features of the Black Sea Region, such as high coastal mountains, as well as the Black Sea itself influence mesoscale atmospheric circulation and cyclone formation in the region. Such effects have been considered for the neighboring Mediterranean Sea. According to [2] there are three kinds of cyclone evolution that can be identified in the Mediterranean: (a) The low-level disturbance does not evolve and remains shallow, weak or moderate and nearly stationary if the upper level potential vorticity (PV) anomaly is absent or too far away to interact with it. In this case real or deep cyclogenesis does not develop. (b) An upper level PV anomaly will create cyclogenesis when arriving over a frontal zone, just under the maximum PV advection at high levels, with or without the presence of a depression at low level. Some Mediterranean cyclogenesis could be, at least partially of this type, comprising those cyclones generated at the quasi-permanent Mediterranean border front or at some internal fronts. (c) The low-level disturbance rapidly deepens and cyclogenesis occurs when the upper level perturbation moves close enough to interact with it. According to [3] about 50% of the Mediterranean cyclones over the sea are located in small areas close to coastal mountains.

Apart form described kinds of cyclogenesis mesoscalse cyclones with similarities to tropical cyclones and polar lows are observed in the Mediterranean Region. These so-called quasitropical cyclones form by means of latent heat release in large cumulus clusters [4,5].

Cyclone climatology in the Black Sea Region is not properly studied. As usual the region is neglected in Mediterranean cyclone studies; only few of them consider both Mediterranean and Black Sea Regions. For instance, in [6, 7] cyclogenesis events in the eastern part of the Black Sea are analyzed and the frequencies of first and minimum central pressure detections throughout the day are plotted for spring and summer. Eastern Black Sea cyclones are short-lived with an average life time of 1 day; they occur, when synoptic upper troughs move over the relatively warm water basin (in winter) or over low-level baroclinity. However these conclusions are based on ECMWF reanalysis with low spatial (1.25°) and temporal (6hr) resolution for a relatively short 17-yr period. The spatial and temporal resolution of ECMWF reanalysis strongly limits the detection of small short-lived cyclones.

A detailed climatology of cyclone behavior and development is necessary for a more complete understanding of the Black Sea Region climate and its variability. The only reason why such study has not been conducted is the lack of data about atmospheric circulation over the sea. However nowadays there are regional models which can simulate local climate using global modeling results.

In present work long-term (1960-1990 yr) dataset with high spatial (25x25km) and temporal (1hr) resolution is used to study mesoscale cyclones over the Black Sea. These data were provided by the regional climate model PRECIS [8] with initial and boundary conditions obtained from the global HadAM3P model. This data does not describe atmospheric evolution that took place in reality, but only one of possible atmospheric evolutions that could take place in the region. On the other hand 30yr period is long enough to assume that simulated climate means are close to real ones. This study is based on PRECIS hourly surface wind fields for 30yr period.

## 1. Methodology

There are a number of methods for cyclone detection and tracking [9]. For instance a cyclone candidate can be identified as local sea level pressure minimum. In present work cyclone identification algorithm is based on Ocubo-Weiss criterion (OWC). The criterion is often used for examining two-dimensional topology structure, in particular, for detection of coherent structures —axisymmetrical vortexes in the wind field [10].

According to Cauchy - Helmholtz theorem, the velocity of any point of an infinitesimal particle is composed of the translational movement velocity of a particle center, the rotation velocity of a particle as an absolutely rigid body with respect to the point, and the velocity related to the particle deformation:

**v**(**r**+d**r**)=**v**(**r**)+**U**·d**r**, where **U** is the tensor with components $U_{ij} = \frac{\partial v_i}{\partial x_j}$, **v**(**r**) – translational velocity.

Tensor antisymmetric part, defined as **A**=(**U**-**U**$^T$)/2, describes rotation, and tensor symmetric part, defined as **S**=(**U**+**U**$^T$)/2, describes pure shear. In two-dimensional case **A** and **S** become

$$\mathbf{A} = \frac{1}{2}\begin{pmatrix} 0 & u_y - v_x \\ v_x - u_y & 0 \end{pmatrix}, \quad \mathbf{S} = \frac{1}{2}\begin{pmatrix} 2u_x & u_y + v_x \\ v_x + u_y & 2v_y \end{pmatrix}.$$

Strain velocity tensor **S** can be decomposed into shear strain tensor **S**$_1$ and irrotational deformation tensor **S**$_2$:

$$\mathbf{S}_1 = \frac{1}{2}\begin{pmatrix} u_x - v_y & u_y + v_x \\ v_x + u_y & v_y - u_x \end{pmatrix}, \quad \mathbf{S}_2 = \frac{1}{2}\begin{pmatrix} u_x + v_y & 0 \\ 0 & u_x + v_y \end{pmatrix}.$$

The quantity $W = 4\det(\mathbf{A} + \mathbf{S}_1) = \omega^2 - s^2$ is so-called Ocubo-Weiss criterion. Here $\omega^2 = (v_x - u_y)^2$ — square vorticity and $s^2 = (u_x - v_y)^2 + (v_x + u_y)^2$ — square strain.

OWC is large and positive for an axisymmetric vortex with solid body rotation. The criterion is minimal and negative for pure shear wind field without rotation. For a shear flow OWC takes on intermediate values. So, OWC criterion allows detection of strong axisymmetric vortexes [10]. Large OWC value corresponds to intense and axisymmetrical vortex.

Cyclone detection method is as follows. A cyclone candidate is identified as a local OWC maximum. To be considered a cyclone this maximum must exceed an empirical threshold. The cyclone tracking algorithm is based on a nearest-neighbor search procedure: a cyclone's trajectory is determined by computing a distance to a cyclone detected in a previous chart. Maximum cyclone velocity is assumed to be 100 km h$^{-1}$. If a cyclone center in the previous chart is within 100 km range from the current one, the previous cyclone is assumed to have moved to a new position. Otherwise it is assumed that the previous cyclone has dissipated and a new cyclone has formed.

Synoptic cyclones are much larger than the Black Sea, so they influence regional atmospheric circulation even when their centers are located over the land. Unfortunately this method allows identifying only those synoptic-scale cyclones whose centers are located over the sea.

In this study only long-lived cyclones, that spent at least 6h hours over the sea, are considered.

## 2. Cyclone statistics

Figure 1a shows the distribution of cyclone centers over the Black Sea for 30 years; figures 1b and 1c show the same distribution but for summer and winter months respectively. Cyclone center is determined as a local OWC maximum. The size of a 'bubble' corresponds to the number of hours spent by cyclones in a 25x25 km cell. As Ocubo-Weiss criterion was calculated by means of central-difference operator; coastal areas were excluded from the analyses. Only cyclones with centers over the Black Sea are analyzed. Synoptic cyclones whose centers moved over the land are not considered.

According to Fig.1a there are three types of cyclonic eddies in the Black Sea Region:
1. Caucasian coastal eddies, whose centers are located near the Caucasian coast
2. Caucasian off-shore eddies, whose centers are located in the southern-eastern part of the sea
3. Ambient cyclonic eddies that originate and move over the whole sea and do not have any particular location.

Now let us consider these three types.

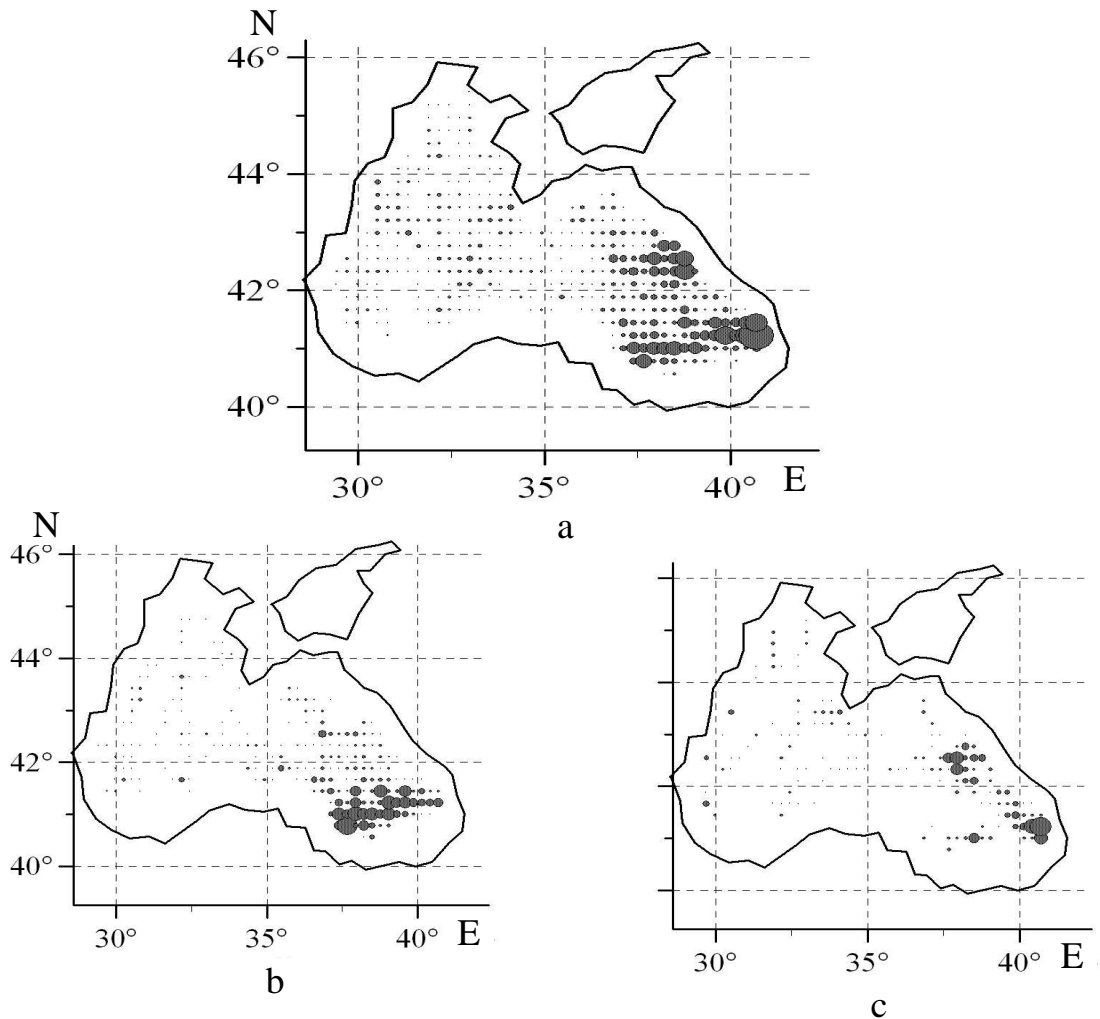

Fig.1. The distribution of cyclone centers over the Black Sea for 30 years. (a) all cyclones, (b) only summer cyclones, (c) only winter cyclones. The size of a «bubble» corresponds to the number of hours spent by cyclones in a 25x25 km cell. Maximum radius of a «bubble» corresponds to 77 hours (a), 31 hours (b) and 23 hours (c) respectively. Minimum radius of a «bubble» corresponds to 1 hour (a)-(c).

*2.1 Caucasian coastal eddies*

The algorithm for Caucasian coastal eddy detection is not based on Ocuba-Weiss criterion. These eddies are localized near the Caucasian coast, so we use a method which involves making a composite wind field. At first, a composite surface wind field is made by averaging wind fields associated with typical Caucasian coastal eddies. Then correlation coefficients between all hourly wind fields and the composite wind field are calculated for the eastern part of the sea. In the end, only wind fields with correlation coefficients above threshold are considered.

Fig. 2 shows a composite surface wind field for Caucasian coastal eddies. According to Fig. 2 such eddy forms when the air flows around the Caucasian mountains. They exist over the southern-eastern part of the sea as a stream with cyclonic vorticity. Actually there is only half of the vortex located over the sea.

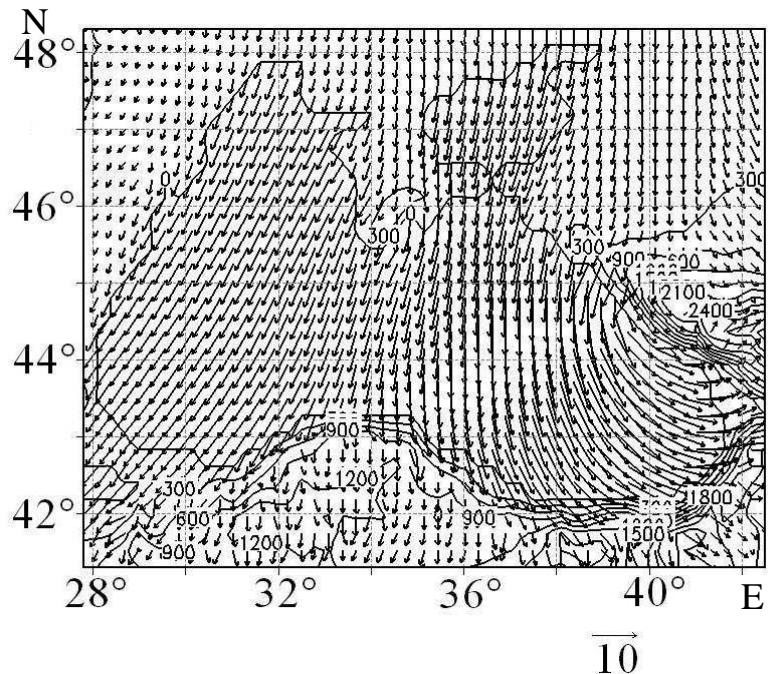

Fig.2. Composite surface wind field (m/s) for Caucasian coastal eddies.

Fig.3a shows histogram of the detected cyclones' maximum intensity. Cyclone's maximum intensity is measured by a maximum value of Ocubo-Weiss criterion. Almost 50% of the detected cyclones spend less then 12 hours over the sea (Fig 3b). Average time spend by Caucasian coastal eddies over the sea equals 15 hours.

Caucasian coastal eddies are characterized by a pronounced seasonal cycle (Fig.4): they appear mostly in late autumn-winter and almost absent in summer. This can be explained by a seasonal variability of general atmospheric circulation in the Black Sea Region. In summer there is a large-scale anticyclonic circulation over Mediterranean and Black Sea Regions [11]. An eastern periphery of the anticyclone as well as north wind in the lower troposphere covers south-western part of Europe. When wind flows around Caucasian mountains cyclonic eddy forms over the eastern part of the sea (Fig. 2)

So, it can be deduced that Caucasian coastal eddies appear mostly in winter when strong north wind flows around Caucasian mountains and exist as a strong south-eastern stream with cyclonic vorticity.

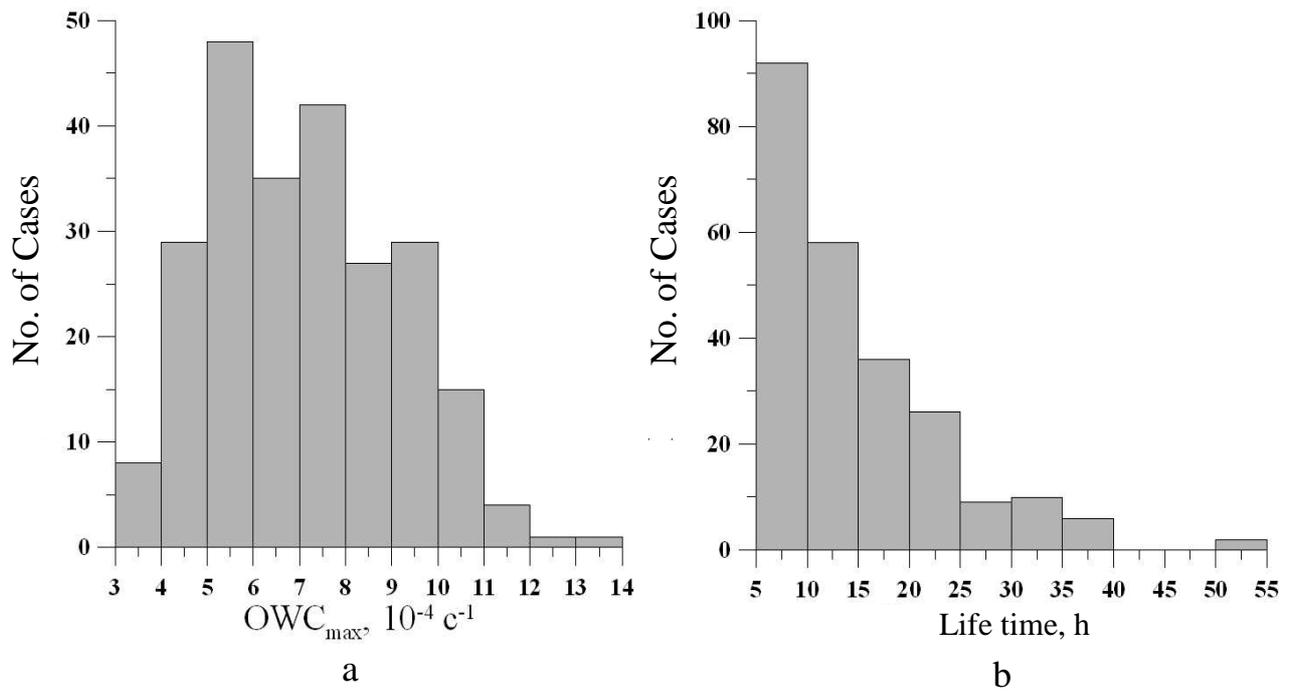

Fig.3. Histograms of Caucasian coastal eddies' characteristics: (a) maximum intensity ($s^{-1}$) and (b) life time (hour).

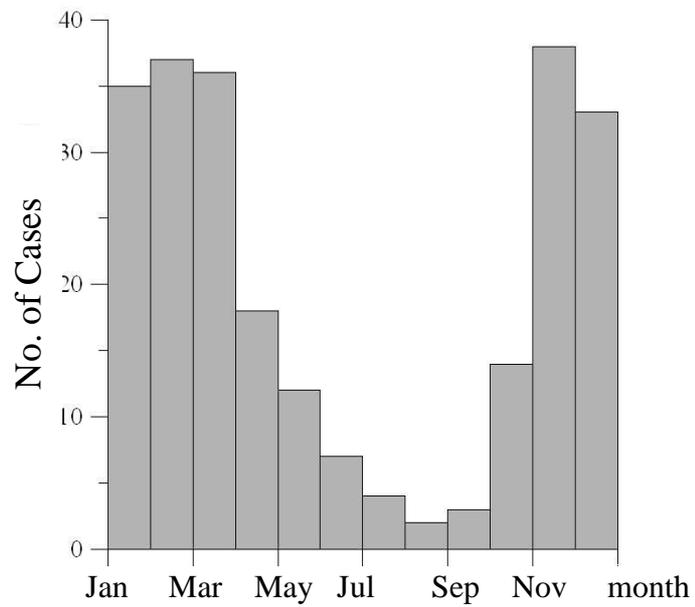

Fig.4 Seasonal cycle of Caucasian coastal eddies.

*2.2 Caucasian off-shore eddies*

The algorithm for Caucasian off-shore eddies detection is the same as for Caucasian coastal eddies. The only difference is that off-shore eddies are more mobile than coastal ones and therefore several composite surface wind fields were made. These composite wind fields differ only by location of an eddy center, i.e. the point of maximum OWC value. As a result of the algorithm 450 off-shore eddies were detected. Fig. 5 shows a composite surface wind field for 90 Caucasian off-shore eddies. It can be regarded as a typical wind field for such eddies.

Fig. 6 shows seasonal cycle of Caucasian off-shore eddies. It differs greatly from that of coastal eddies. These eddies appear mostly in summer - early autumn. Also there is a pronounced diurnal cycle: they appear in the evening, start to intensify at night and dissipate in the morning next day.

So Caucasian off-shore eddies are short-lived, form mostly in summer and exist over the southern-eastern part of the Black Sea as an axisymmetrical surface eddy. Also they are characterized by a pronounced diurnal cycle.

Physical nature of these eddies is yet to be investigated. For the time being it can supposed that the Caucasian off-shore eddies have much in common with Catalina eddies which form off the coast of southern California [12,13]. In particular, both Caucasian and Catalina eddies have diurnal cycle.

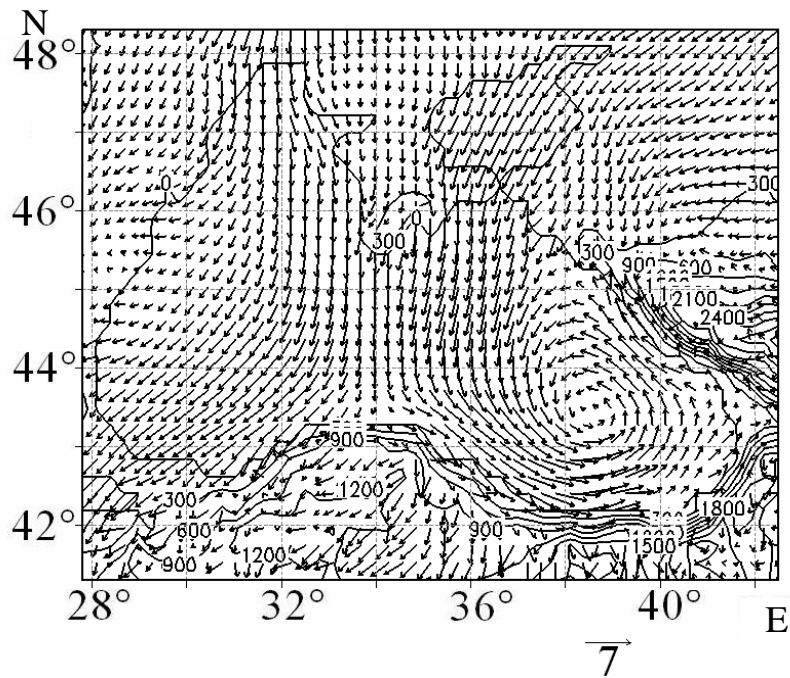

Fig.5. Composite surface wind field (m/s) for Caucasian off-shore eddies.

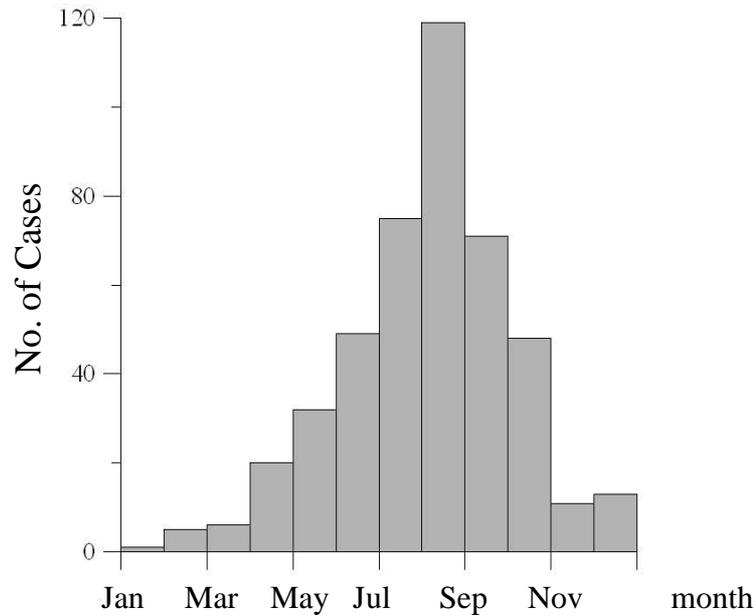

Fig.6. Seasonal cycle of Caucasian off-shore eddies.

*2.3 Cyclonic eddies*

Ocubo –Weiss criterion is used to detect cyclonic eddies that do not originate near the Caucasian mountains. These cyclones have various spatial and temporal scales: from mesoscale eddies with spatial scale of 100 km and temporal scale of 1 day to common synoptic cyclones moving over the sea.

Mesoscale cyclonic eddies usually appear in the western part of the sea and move in a north-easterly direction sometimes around the Crimean peninsula and make landfall. Apart from this typical track there are other various tracks.

In general, centers of ambient cyclones are uniformly distributed over the sea. They move occasionally but mostly in a north/ north-easterly direction. Figure 9a shows histogram of the detected cyclones' maximum intensity; average value is about 5.8 $s^{-1}$. Figure 9b shows histogram of the detected cyclones' life time. Almost 80% of the detected cyclones spend less then 12 hours over the sea and 20 hours in average; only few of them spend 2-3 days. For north-easterly tracks it means that a cyclone moves with velocity of 20-30 m/s over the sea. In average there was about 1 cyclone over the Black Sea in 2-3 months. Mind that we consider only those synoptic cyclones whose centers are located over the sea. In general the number of synoptic cyclones that influence the regional circulation is much larger.

As it was expected the most intense cyclones are at the same time the most long-lived: correlation coefficient between intensity and life time equals 0.7. It was detected about 130 cyclones over the sea during 30 years. Of course the number of detected cyclones depends on threshold. In present work threshold of $3.8 \cdot 10^{-4}$ $s^{-1}$ was used which corresponds to large surface wind velocities, more then 10 m/s.

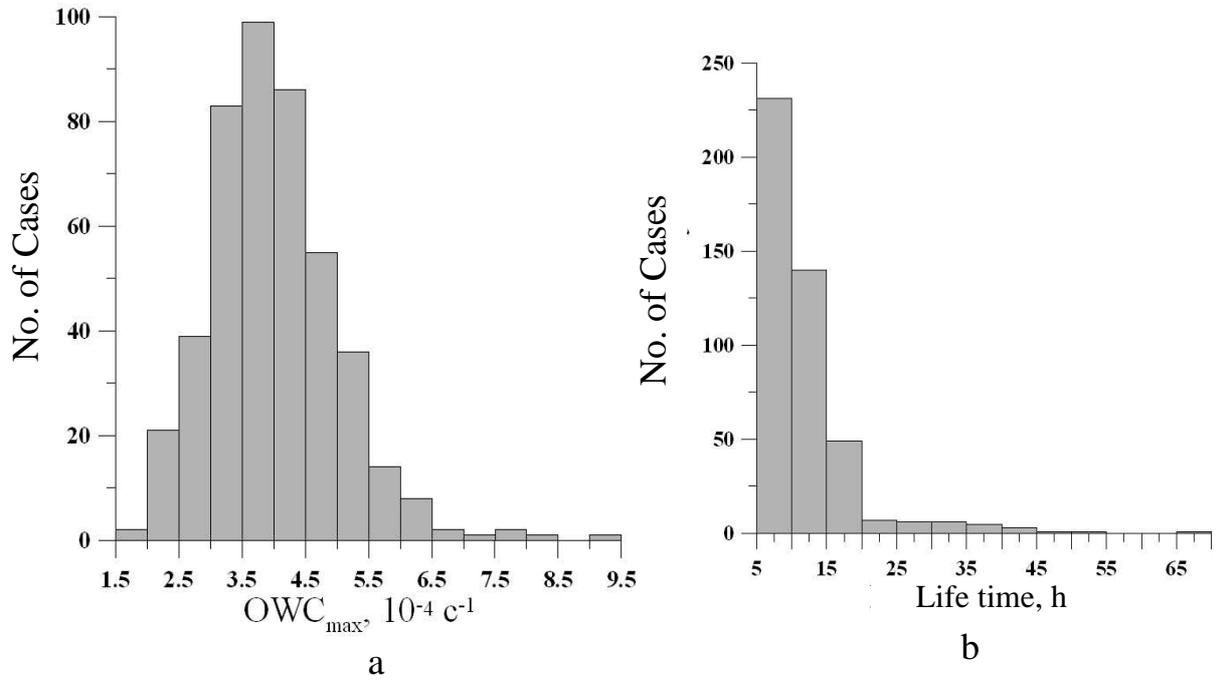

Fig.7. Histograms of the Caucasian off-shore eddies' characteristics: (a) maximum intensity ($s^{-1}$) and (b) life time (hour).

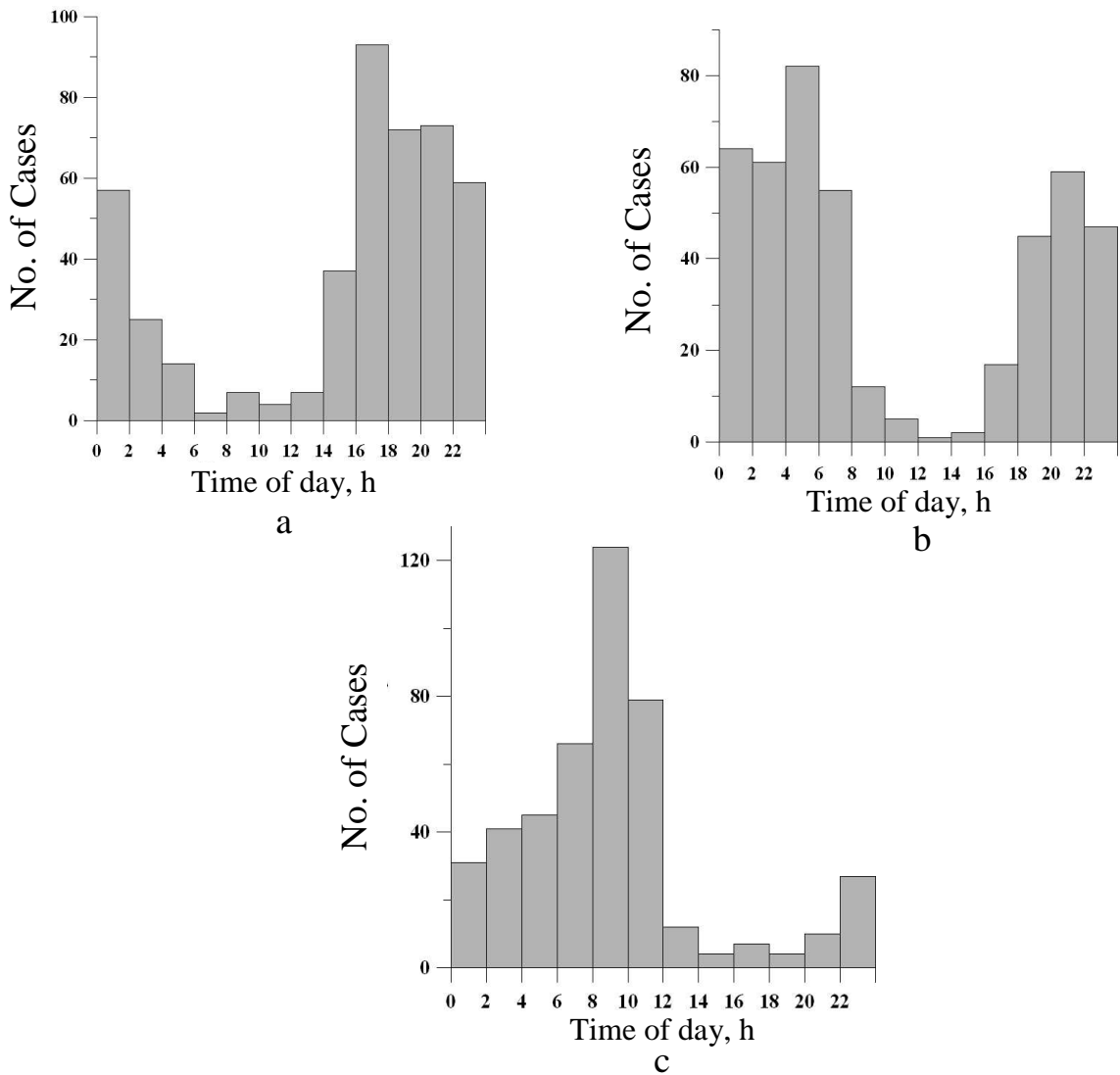

Fig.8 Diurnal cycle of Caucasian off-shore eddies: (a) origin, (b) maximum intensity, (c) dissipation.

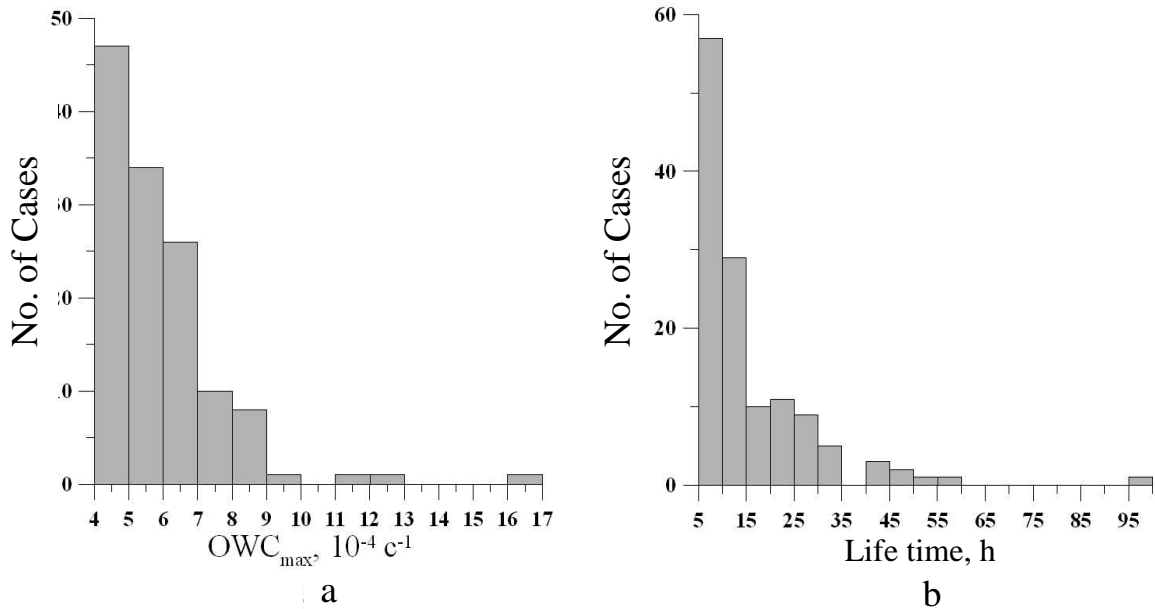

Fig.9. Histograms of ambient cyclones' characteristics: (a) maximum intensity ($s^{-1}$) and (b) life time (hour).

*2.4 Quasitropical cyclones*

Some ambient cyclones can be regarded as quasitropical ones. They occur only if certain conditions are fulfilled, such as convective instability of marine troposphere, small horizontal wind speed and others [14, 15]. One of such cyclones formed in September 2005 and was thoroughly studied in [14]. Its structure was similar to that of a tropical cyclone and surface wind speed was anomalously large (27 ms$^{-1}$). Unfortunately in this study only surface wind data is available which does not allow identifying quasitropical cyclones with all confidence. We can only identify candidate cyclones that could be qusitropical ones. These nearly axisymmetrical cyclones originated and moved over the sea. Also they developed in autumn the most favorable season for quasitropical cyclone formation [14]. There were three mesoscale axisymmetric cyclones with wind speed over 25 ms$^{-1}$ and life time 2-5 days detected in 30-yr dataset.

**Conclusions**

A 30-yr dataset was analyzed and four cyclone types in the Black Sea were determined: Caucasian coastal eddies, Caucasian offshore eddies, ambient eddies and quastropical eddies.

Caucasian coastal eddies appear mostly in winter when strong north wind flows around Caucasian mountains and exist as a strong south-eastern stream with cyclonic vorticity.

Caucasian off-shore eddies are short-lived, form mostly in summer and exist over the southern-eastern part of the Black Sea as an axisymmetrical surface eddy. They are also characterized by a pronounced diurnal cycle. The Caucasian off-shore eddies have much in common with Catalina eddies which form off the coast of southern California.

Mesoscale cyclonic eddies have various spatial and temporal scales from hundred kilometers and a day to typical parameters of synoptic cyclones that move over the sea. Centers of ambient cyclones are scattered over the whole Black Sea basin. They move occasionally but mostly in a north/ north-easterly direction. Ambient cyclones vary from common large scale extratropical cyclones to secondary mesoscale eddies associated with coastal mountains influence and barotrophic/baloclinic instability of marine atmosphere. These cyclones will be further systemized with the help of high-resolution regional analysis.